\documentclass[aps,prd,onecolumn,superscriptaddress,showpacs,nofootinbib,floatfix,onecolumn]{revtex4}

\usepackage{dcolumn}
\usepackage{bm}

\usepackage[dvips]{graphicx}
\usepackage{float}
\usepackage{epsfig,subfigure}
\usepackage{longtable}
\usepackage{rotating}
\usepackage{amsmath}
\usepackage{amsfonts}
\usepackage{amssymb}
\usepackage{dcolumn}
\usepackage{hyperref}   
\usepackage{axodraw4j}
\usepackage{multirow}

\usepackage{color}
\usepackage{txfonts}

\newcommand{\sqrts}{\sqrt{s}}
\newcommand{\sqrtsnn}{\sqrt{s_{_{\mbox{\rm \tiny{NN}}}}}}
\newcommand{\alphaS}{\alpha_{\rm s}}

\newcommand{\pp}{{\rm{p-p}}}

\newcommand{\ppbar}{{\rm{p-$\bar{\rm p}$}}}
\newcommand{\pA}{{\rm{p-A}}}
\newcommand{\AaAa}{{\rm{A-A}}}

\newcommand{\NN}{{\rm{N-N}}}

\newcommand{\PbPb}{{\rm{Pb-Pb}}}

\newcommand{\dtwor}{{\rm{d^2r}}}

\newcommand{\sigmaDPS}{\sigma^{{\rm {\tiny DPS}}}}
\newcommand{\sigmaDPSjpsijpsi}{\sigma^{{\rm {\tiny DPS}}}_{{\jpsi\jpsi}}}
\newcommand{\sigmaDPSone}{\sigma^{{\rm {\tiny DPS,1}}}}
\newcommand{\sigmaDPStwo}{\sigma^{{\rm {\tiny DPS,2}}}}
\newcommand{\sigmaDPSthree}{\sigma^{{\rm {\tiny DPS,3}}}}
\newcommand{\sigmaSPS}{\sigma^{{\rm {\tiny SPS}}}}

\newcommand{\sigmaeffpp}{\sigma_{_{\rm eff,pp}}}

\newcommand{\sigmaeffAA}{\sigma_{_{\rm eff,AA}}}

\newcommand{\QQbar}    {\rm{Q$\overline{\rm Q}$}}
\newcommand{\ccbar}    {\rm{c$\bar{\rm c}$}}
\newcommand{\bbbar}    {\rm{b$\bar{\rm b}$}}
\newcommand{\jpsi}{J/\psi}

\newcommand{\ups}{\Upsilon}

\begin{document}


\title{Enhanced $\jpsi$-pair production from double parton scatterings in \\ nucleus-nucleus collisions at the Large Hadron Collider}
 
\author{David~d'Enterria}
\affiliation{CERN, PH Department, 1211 Geneva, Switzerland}

\author{Alexander~M.~Snigirev}
\affiliation{Skobeltsyn Institute of Nuclear Physics, Moscow State University, 119991 Moscow, Russia}


\begin{abstract}
\noindent
A generic expression of double-parton scattering cross sections in high-energy nucleus-nucleus (\AaAa)
collisions is derived as a function of the corresponding single-parton hard cross sections and of the \AaAa\
event centrality. We consider the case of prompt-$\jpsi$ production in lead-lead (\PbPb)
at the CERN Large Hadron Collider and find that about 20\% (35\%) of the $\jpsi$ events in
minimum-bias (most central) collisions contain a second $\jpsi$ from double parton interactions. In \PbPb\ at
5.5~TeV, in the absence of final-state effects, about 240 double-$\jpsi$ events are expected per unit
midrapidity and per inverse-nanobarn in the dilepton decay modes. The implications of double-$\jpsi$
production on the interpretation of the observed $\jpsi$ suppression in \AaAa\ collisions are discussed.
\end{abstract}




\maketitle

\paragraph*{\bf Introduction --}
\label{sec:1}

The production of heavy-quark bound states of the charmonium ($\jpsi$) and bottomonium ($\ups$) families 
in high-energy proton-proton (\pp) and  nucleus-nucleus (\AaAa) collisions is governed by both perturbative and
non-perturbative aspects of Quantum Chromodynamics (QCD) and has been extensively studied at fixed-target and
collider energies~\cite{Brambilla:2010cs}. For the most part, a pair of charm or bottom quarks (\ccbar,
\bbbar) is first produced in a hard gluon-gluon collision with 
cross sections computable via perturbative QCD (pQCD) calculations. The subsequent evolution of the
\QQbar\ pair towards a color-singlet bound state is a non-perturbative process described in various
theoretical approaches including color-singlet and color-octet mechanisms, non-relativistic QCD effective
field theory, or color evaporation models (see e.g.~\cite{Lansberg:2006dh} for a review).\\ 

In the case of \AaAa\ collisions, quarkonium has been proposed as a key probe of the
thermodynamical properties of the hot QCD medium produced in the course of the collision~\cite{matsui_satz}. 
Analysis of quarkonia correlators and potentials in finite-temperature lattice QCD~\cite{Datta:2003ww} 
indicate that the different \ccbar\ and \bbbar\ bound-states dissociate at temperatures $T$ for which 
the color (Debye) screening radius of the medium falls below their corresponding \QQbar\ binding radius.
Experimental confirmation of such a quarkonia dissociation pattern should provide a 
direct means to determine the temperature of the produced quark-gluon-plasma (QGP)~\cite{Karsch:2005nk}.
Surprisingly, $\jpsi$ production in lead-lead (\PbPb) collisions at the 
LHC~\cite{Aad:2010aa,Chatrchyan:2012np,Abelev:2012rv,Suire:2012gt} is observed to be less suppressed 
-- compared to baseline \pp\ collisions at the same energy --
than at the Relativistic Heavy-Ion Collider (RHIC)~\cite{Adare:2006ns} 
despite the fact that the average medium temperature at LHC nucleon-nucleon center-of-mass (c.m.)
energies ($\sqrtsnn$~=~2.76~TeV) is at least 30\% higher than at RHIC ($\sqrtsnn$~=~200~GeV)~\cite{Chatrchyan:2012mb}.
Approaches combining $\jpsi$ dissociation in a deconfined phase plus regeneration due to charm-quark
recombination~\cite{Andronic:2007bi,Zhao:2011cv} can reproduce the observed trends in the data 
although the model parameters ($\sigma_{c\bar c}$ cross section, medium density, ...) 
need to be validated with other LHC observations.\\ 

In this paper we discuss and quantify for the first time in the literature the role of
double-parton-scattering (DPS) processes in ultrarelativistic heavy-ion collisions, considering specifically
the case of double-$\jpsi$ production in \PbPb\ at LHC energies. Due to the fast increase
of the parton flux at small parton fractional momenta, $x$~$\equiv$~$p_{\rm parton}/p_{\rm hadron}$, 
the probability of having multiple hard parton interactions (MPI) occurring simultaneously at
different impact parameters increases rapidly with collision energy and constitutes
a significant source of particle production at semihard scales of a few~GeV in \pp\ and, in particular, 
\AaAa\ collisions~\cite{MPI}. 
The evidence for DPS processes producing two independently-identified hard particles in the same collision is
currently based on \pp\ and \ppbar\ measurements of final-states containing multi-jets, and jets plus
photons~\cite{Abe:1997bp,Abazov:2009gc} or W$^\pm$ bosons~\cite{Aad:2013bjm}
showing an excess of events in various differential distributions with respect to the expectations
from contributions from single-parton scatterings (SPS) alone.
LHC \pp\ measurements of double-$\jpsi$ production~\cite{Aaij:2011yc} as well as of single-$\jpsi$ production
as a function of the event multiplicity~\cite{Abelev:2012rz} have been also interpreted 
in the context of DPS~\cite{Kom:2011bd,Baranov:2011ch,Novoselov:2011ff,Baranov:2012re} and MPI models respectively.\\

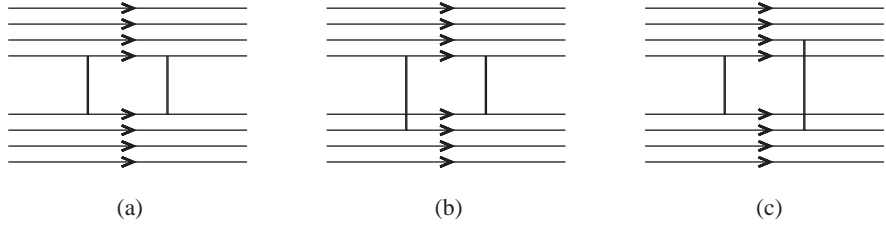
\begin{figure}
\begin{minipage}{0.99\textwidth}
\begin{picture}(100,120)
\Line[arrow,arrowinset=0.9](0,20)(90,20)
\Line[arrow,arrowinset=0.9](0,26)(90,26)
\Line[arrow,arrowinset=0.9](0,32)(90,32)
\Line[arrow,arrowinset=0.9](0,38)(90,38)
\Line[arrow,arrowinset=0.9](0,60)(90,60)
\Line[arrow,arrowinset=0.9](0,66)(90,66)
\Line[arrow,arrowinset=0.9](0,72)(90,72)
\Line[arrow,arrowinset=0.9](0,78)(90,78)
\Line[double,sep=0.4](30,38)(30,60)
\Line[double,sep=0.4](60,38)(60,60)

\Line[arrow,arrowinset=0.9](120,20)(210,20)
\Line[arrow,arrowinset=0.9](120,26)(210,26)
\Line[arrow,arrowinset=0.9](120,32)(210,32)
\Line[arrow,arrowinset=0.9](120,38)(210,38)
\Line[arrow,arrowinset=0.9](120,60)(210,60)
\Line[arrow,arrowinset=0.9](120,66)(210,66)
\Line[arrow,arrowinset=0.9](120,72)(210,72)
\Line[arrow,arrowinset=0.9](120,78)(210,78)
\Line[double,sep=0.4](150,32)(150,60)
\Line[double,sep=0.4](180,38)(180,60)


\Line[arrow,arrowinset=0.9](240,20)(330,20)
\Line[arrow,arrowinset=0.9](240,26)(330,26)
\Line[arrow,arrowinset=0.9](240,32)(330,32)
\Line[arrow,arrowinset=0.9](240,38)(330,38)
\Line[arrow,arrowinset=0.9](240,60)(330,60)
\Line[arrow,arrowinset=0.9](240,66)(330,66)
\Line[arrow,arrowinset=0.9](240,72)(330,72)
\Line[arrow,arrowinset=0.9](240,78)(330,78)
\Line[double,sep=0.4](270,38)(270,60)
\Line[double,sep=0.4](300,32)(300,66)

\end{picture}
\hspace{-2.15cm}(a)\hspace{3.8cm} (b)\hspace{3.9cm}(c)\\

\caption{Schematic DPS contributions in \AaAa\ collisions: 
(a) The two colliding partons belong to the same pair of nucleons, 
(b) partons from one nucleon in one nucleus collide with partons from two different nucleons in the other nucleus, and 
(c) the two colliding partons belong to two different nucleons from both nuclei.
\label{fig:diags}}
\end{minipage}
\end{figure}

We investigate DPS in \AaAa\ collisions following a similar study for \pA\
collisions~\cite{d'Enterria:2012qx}, extending it to also include the centrality-dependence of the DPS cross
sections. The larger transverse parton density in nuclei compared to protons results in enhanced
\AaAa\ DPS contributions coming from interactions where the two partons belong or not to the same pair of
nucleons of the colliding nuclei (Fig.~\ref{fig:diags}). 
Consequently, in \PbPb\ at LHC energies we expect a non-negligible probability of two parton-parton
interactions independently producing two $\jpsi$ mesons in the same nuclear collision.\\


\paragraph*{\bf Cross sections for double parton scattering in proton and nuclear collisions --}
\label{sec:2}
The DPS cross section in \pp\ collisions can be theoretically computed from the convolution of parton
distribution functions (PDF) and elementary cross sections summed over all involved partons (see e.g.~\cite{Diehl:2011yj})
\begin{eqnarray} \label{eq:DPS_generic}
\sigmaDPS_{(pp\to a b)} = & & \left(\frac{m}{2}\right) \sum \limits_{i,j,k,l} \int \Gamma_{p}^{ij}(x_1, x_2; {\bf b_1},{\bf b_2}; Q^2_1, Q^2_2)\nonumber\\
& &\times\,\hat{\sigma}^{ik}_{a}(x_1, x_1',Q^2_1) \, \hat{\sigma}^{jl}_{b}(x_2, x_2',Q^2_2)\\ 
& &\times\,\Gamma_{p}^{kl}(x_1', x_2'; {\bf b_1} - {\bf b},{\bf b_2} - {\bf b}; Q^2_1, Q^2_2)\, dx_1 dx_2 dx_1' dx_2' d^2b_1 d^2b_2 d^2b\,,\nonumber
\end{eqnarray}
where $\Gamma_{p}^{ij}(x_1, x_2;{\bf b_1},{\bf b_2}; Q^2_1, Q^2_2)$ are double-PDF 
which depend on the longitudinal momentum fractions $x_1$ and $x_2$ and transverse positions
${\bf b_1}$ and ${\bf b_2}$ of the two partons undergoing the hard processes 
at scales $Q_1$ and $Q_2$, $\hat{\sigma}^{ik}_{a}$ and $\hat{\sigma}^{jl}_{b}$ are the
parton-level subprocess cross sections, and ${\bf b}$ is the impact parameter vector 
connecting the centers of the colliding protons in the trans\-verse plane. The combinatorial factor $m/2$
accounts for indistinguishable ($m=1$) and distinguishable ($m=2$) final-states.
In a  model-independent way, the cross section of double parton
scattering can be expressed in the simple generic form
\begin{equation} 
\sigmaDPS_{(pp\to a b)} = \left(\frac{m}{2}\right) \frac{\sigmaSPS_{(pp\to a)} \cdot \sigmaSPS_{(pp\to b)}}{\sigmaeffpp}\,,
\label{eq:doubleAB}
\end{equation}
where $\sigmaSPS$ is the inclusive single-hard scattering cross section, computable perturbatively to
a given order in $\alphaS$,
\begin{eqnarray} 
\sigmaSPS_{(pp \to a)} & = & \sum \limits_{i,k} \int D^{i}_{p}(x_1; Q^2_1) f({\bf b_1})\,
\hat{\sigma}^{ik}_{a}(x_1, x_1') \times\, D^{k}_{p}(x_1'; Q^2_1)f( {\bf b_1} - {\bf b}) dx_1 dx_1' d^2b_1
d^2b\nonumber \\
& = & \sum \limits_{i,k} \int D^{i}_{p}(x_1; Q^2_1) \,\hat{\sigma}^{ik}_{a}(x_1, x_1') \,D^{k}_{p}(x_1'; Q^2_1) dx_1 dx_1'\,,
\label{eq:hardS}
\end{eqnarray}
and $\sigmaeffpp$ is a normalization cross section that encodes all ``DPS unknowns'' into a single parameter
which can be experimentally measured. A simple relationship between Eqs.~(\ref{eq:DPS_generic})
and~(\ref{eq:doubleAB}) can be obtained by (i) decomposing the double PDF into longitudinal and transverse
components, with the latter expressed in terms of the overlap function $t({\bf b}) = \int
f({\bf b_1}) f({\bf b_1 -b})d^2b_1 $ for a given parton transverse thickness function $f({\bf b})$ 
representing the effective transverse overlap area of partonic interactions that produce the DPS process, 
and (ii) making the assumption that the longitudinal component reduces to the ``diagonal'' product of two
independent single-PDF, $D^{i}_{p}(x_1; Q^2_1)$. 
Under such simplifying approximations one can identify $\sigmaeffpp$ with the inverse of the
nuclear-overlap function squared:
$\sigmaeffpp = \left[ \int d^2b \, t^2({\bf b})\right]^{-1}$,
whose numerical value, $\sigmaeffpp \approx$~14~mb, 
has been obtained empirically from fits to \pp\ and \ppbar\ 
data~\cite{Abe:1997bp,Abazov:2009gc,Aad:2013bjm}.\\

To compute the DPS cross section in nucleus-nucleus collisions we proceed as done for \pA\
in~\cite{d'Enterria:2012qx}. The parton flux is enhanced by the number $A$ of nucleons in each
nucleus and the single-parton cross section is simply expected to be that of \pp\ -- or, more exactly,
nucleon-nucleon (\NN) collisions taking into account shadowing effects in the nuclear PDF (see below) --
scaled by the factor $A^2$, i.e.
\begin{eqnarray} 
\sigmaSPS_{(AA \to a)} = \sigmaSPS_{(NN \to a)} \int \rm T_{\rm A}({\bf b_1}) \rm T_{\rm A}({\bf b_1-b}) d^2b_1 d^2b= \sigmaSPS_{(NN \to a)} \int \rm T_{\rm AA}({\bf b})d^2b =A^2 \cdot \sigmaSPS_{(NN \to a)}\,.
\label{eq:sigmaSPSAA}
\end{eqnarray}
Here $T_{\rm A}({\bf b})$ is the nuclear thickness function at impact parameter vector
${\bf b}$ connecting the centers of the colliding nucleus in the transverse plane, and
$\rm{T}_{AA}({\bf b})$ the standard nuclear overlap function normalised to $A^2$~\cite{d'Enterria:2003qs}. 
The DPS ${\rm A}$-${\rm A}$ cross section is thus the sum of three terms,
corresponding to the diagrams of Fig.~(\ref{fig:diags}):
\begin{enumerate}
\item The first term corresponding to Fig.~\ref{fig:diags}(a) is just, similarly to the SPS cross sections
Eq.~(\ref{eq:sigmaSPSAA}), the DPS cross section in \NN\ collisions scaled by $A^2$
\begin{eqnarray} 
\sigmaDPSone_{(AA \to a b)} = A^2 \cdot \sigmaDPS_{(NN \to a b)}\,.
\label{eq:AAdoubleAB1}
\end{eqnarray} 
\item The second term, Fig.~\ref{fig:diags}(b), accounts for interactions with partons from one nucleon
in one nucleus with partons from two different nucleons in the other nucleus. This term was originally
derived in~\cite{Strikman:2001gz} in the context of \pA\ collisions,  
\begin{eqnarray} 
\label{eq:AAdoubleAB2}
\sigmaDPStwo_{(AA \to a b)} = 2\sigmaDPS_{(NN \to a b)} \cdot \sigmaeffpp \cdot \rm{T}_{2,AA}\,,
\end{eqnarray}
with
\begin{eqnarray}
\rm{T}_{2,AA} = \frac{A-1}{A} \int \rm T_{\rm A}({\bf b_1}) \rm T_{\rm A}({\bf b_1-b}) \rm T_{\rm A}({\bf
  b_1-b}) d^2b_1 d^2b = (A-1)\int \dtwor\, T_{\rm A}^2({\bf r}) = (A-1) \cdot \rm{T}_{AA}(0).
\label{eq:AATpAsq}
\end{eqnarray}
\item The third contribution from interactions of partons from two different nucleon in one nucleus with 
partons from two different nucleons in the other nucleus, Fig.~\ref{fig:diags}(c),
reads
\begin{eqnarray} 
\label{eq:AAdoubleAB3}
\sigmaDPSthree_{(AA \to a b)} = \sigmaDPS_{(NN \to a b)} \cdot \sigmaeffpp \cdot \rm{T}_{3,AA}\,,
\end{eqnarray}
with
\begin{eqnarray}
\rm{T}_{3,AA} = \left(\frac{A-1}{A}\right)^2 \int \rm T_{\rm A}({\bf b_1}) \rm T_{\rm A}({\bf b_2}) \rm T_{\rm A}({\bf
  b_1-b}) \rm T_{\rm A}({\bf b_2-b}) d^2b_1 d^2b_2 d^2b 
= \left(\frac{A-1}{A}\right)^2 \int \dtwor\, T_{\rm AA}^2({\bf r}) .
\label{eq:AATpAsq3}
\end{eqnarray}
where the integral of the nuclear overlap function squared does not depend much on the precise shape of
the transverse parton density in the nucleus, amounting to $A^2/1.94\cdot\rm{T}_{AA}(0)$ for a hard-sphere
and $A^2/2\cdot\rm{T}_{AA}(0)$ for a Gaussian profile.
\end{enumerate}
The factors $(A-1)/A$ and $[(A-1)/A]^2$ in the two last terms 
take into account the difference between the number of nucleon pairs and the number of {\it different} nucleon pairs. 
Adding (\ref{eq:AAdoubleAB1}), (\ref{eq:AAdoubleAB2}) and (\ref{eq:AAdoubleAB3}), 
the inclusive cross section of a DPS process with two hard parton subprocesses $a$ and $b$ in \AaAa\
collisions (with A large, so that A~$-~1~\approx$~A) can be written as  
\begin{eqnarray} 
\sigmaDPS_{(AA\to a b)} & =& A^2 \,\sigmaDPS_{(NN \to a b)}\cdot 
\left[1+\frac{2\,(A-1)}{A^2}\,\sigmaeffpp \,\int \dtwor\, \rm T_{\rm A}^2({\bf r})\,+\,\left(\frac{A-1}{A^2}\right)^2\, \sigmaeffpp \,\rm\int \dtwor\, \rm{T}_{\rm AA}^2({\bf r})\right] \label{eq:doubleAA1}\\
& \approx & A^2 \,\sigmaDPS_{(NN \to a b)}\cdot \left[1+\frac{2}{A}\,\sigmaeffpp \,\rm{T}_{AA}(0)\,+\,\frac{1}{2}\,
  \sigmaeffpp \, \rm{T}_{\rm AA}(0)\right] \,,
\label{eq:doubleAA}
\end{eqnarray}
where the term in parentheses follows a dependence of the type $A^{4/3}/6$ and thus
$\sigmaDPS$ in \AaAa\ increases roughly as $A^{3.3}/5$ compared to its value in \pp\ collisions.
The DPS cross sections in \AaAa\ are practically unaffected by the value of $\sigmaeffpp$ but dominated
instead by double-parton interactions from different nucleons in both nuclei, and thus less sensitive to
possible extra ``non-diagonal'' parton interference terms~\cite{Strikman:2001gz}, computed for light
nuclei in~\cite{Treleani:2012zi}.\\

For $^{208}$Pb\,-$^{208}$Pb collisions, in the simplest hard-sphere approximation for a 
uniform density of radius $R_A=r_0 A^{1/3}$ and $r_0=1.25$~fm, the nuclear overlap function 
at $b$~=~0 is $\rm{T}_{AA}(0) = 9 A^2/(8 \pi R_A^2)$~=~31.5~mb$^{-1}$.
A direct evaluation of the integral using the measured Fermi-Dirac spatial density for the Pb nucleus
($R_A$~=~6.624~fm and surface thickness $a$~=~0.546~fm)~\cite{deJager} 
yields $\rm{T}_{AA}(0)$~=~30.4~mb$^{-1}$. 
Using the latter $\rm{T}_{AA}(0)$ value and $\sigmaeffpp$~=~14~mb, the
expression in parentheses in Eq.~(\ref{eq:doubleAA}) -- which quantifies the total DPS enhancement
factor in \AaAa\ compared to \NN\ collisions, Eq.~(\ref{eq:AAdoubleAB1}) -- is found to be of the order of 200, 
dominated by the hard double nucleon scattering contributions, Fig.~\ref{fig:diags}(c). 
The final DPS cross section ``pocket formula'' in nucleus-nucleus collisions 
can be obtained combining Eqs.~(\ref{eq:doubleAB}) and (\ref{eq:doubleAA}):
\begin{eqnarray} 
\sigmaDPS_{(AA\to a b)} = \left(\frac{m}{2}\right) \frac{\sigmaSPS_{(NN \to a)} \cdot \sigmaSPS_{(NN \to b)}}{\sigmaeffAA},
\label{eq:sigmaAADPS}
\end{eqnarray}
with the effective \AaAa\ normalization cross section for \PbPb\ amounting to
\begin{eqnarray} 
\sigma_{\rm eff,AA} = \frac{1}{A^2\left[\sigmaeffpp^{-1}+\frac{2}{A}\,\rm{T}_{AA}(0)\,+\,\frac{1}{2}\,\rm{T}_{\rm AA}(0)\right]} = 1.5 \mbox{ nb} \,.
\label{eq:sigmaeffAA}
\end{eqnarray}
The relative contributions of the three terms in the denominator, corresponding to the diagrams of
Fig.~\ref{fig:diags}, are approximately 1:4:200. We note that Eq.~(\ref{eq:sigmaeffAA}) is valid only for pQCD
processes with cross sections $\sigmaSPS_{(NN \to a)}$ smaller than about $A^2\,\sigma_{\rm eff,AA}$ (which
holds for the $\jpsi$ case of interest here), otherwise one would need to reinterpret it to account for
triple (and higher multiplicity) parton scatterings. 
Numerically we see that whereas the single-parton cross sections in \PbPb\ collisions,
Eq.~(\ref{eq:sigmaSPSAA}), are enhanced by a factor of $A^2~\simeq~4\cdot 10^4$ compared to that in \pp\
collisions, the corresponding double-parton cross sections are enhanced by a much higher factor of 
$\sigmaeffpp\,/\sigma_{\rm eff,AA}\propto A^{3.3}/5 \simeq 9 \cdot 10^6$.\\

\paragraph*{\bf Centrality dependence of the DPS cross sections  --}
\label{sec:3}
The cross sections discussed so far are for ``minimum bias'' (MB) \AaAa\ collisions without any selection in
the reaction centrality. The cross sections for single and double-parton scattering within an interval of impact
parameters [b$_1$,b$_2$], corresponding to a given centrality percentile, f$_{\%}=$~0--100\%, of the total
\AaAa\ cross section $\sigma_{\rm AA}$, with average nuclear overlap function $\langle T_{\rm AA}[b_1,b_2]\rangle$ are
\begin{eqnarray} 
\sigmaSPS_{(AA \to a)}[b_1,b_2] &=& A^2 \cdot \sigmaSPS_{(NN \to a)}\cdot f_1[b_1,b_2] = 
\sigmaSPS_{(NN \to a)}\cdot f_{\%}\,\sigma_{\rm AA} \cdot \langle T_{\rm AA}[b_1,b_2]\rangle,\label{eq:singleAA_b}\\
\sigmaDPS_{(AA\to a b)}[b_1,b_2] &=& A^2 \cdot \sigmaDPS_{(NN \to a b)}\cdot \rm f_1[b_1,b_2]
\times \left[1+\frac{2}{A}\,\sigmaeffpp \,\rm T_{\rm AA}({0})\,\frac{f_2[b_1,b_2]}{f_1[b_1,b_2]} + \sigmaeffpp \,\rm T_{\rm AA}({0})\, \frac{f_3[b_1,b_2]}{f_1[b_1,b_2]}\right], \label{eq:doubleAA_b}
\end{eqnarray}
where the latter has been obtained integrating Eq.~(\ref{eq:doubleAA1}) over b$_1~<$~b~$<$~b$_2$ and 
where the three dimensionless and appropriately-normalized fractions f$_1$, f$_2$, and f$_3$
read
\begin{eqnarray} 
f_1[b_1,b_2] &=& \frac{2\pi}{A^2} \int_{b_1}^{b_2}bdb \,\rm T_{\rm AA}({b}) = \frac{f_{\%}\,\sigma_{\rm AA}}{A^2}\,\rm
\langle T_{\rm AA}[b_1,b_2]\rangle,\nonumber\\
f_2[b_1,b_2] &=& \frac{2\pi}{A\, \rm T_{\rm AA}({0}) }\int_{b_1}^{b_2}bdb \int d^2b_1\, \rm T_{\rm A}({\bf b_1}) \rm T_{\rm A}({\bf b_1-b}) \rm T_{\rm A}({\bf b_1-b}),\nonumber\\
f_3[b_1,b_2] &=& \frac{2\pi}{A^2\, \rm T_{\rm AA}({0}) } \int_{b_1}^{b_2}bdb\, \rm T_{\rm AA}^2({b}).\nonumber
\label{eq:f3}
\end{eqnarray}
We can evaluate the integrals f$_2$, and f$_3$ for small enough centrality bins around a given impact parameter $b$.
The dominant f$_3$/f$_1$ contribution in Eq.~(\ref{eq:doubleAA_b}) is simply given by the ratio 
$\langle\rm T_{\rm AA}[b_1,b_2]\rangle/\rm T_{\rm AA}({0})$ which is practically insensitive (except for very
peripheral collisions) to the precise shape of the nucleon density in the nucleus~\cite{Lokhtin:2000wm}. 
The second centrality-dependent DPS term, f$_2$/f$_1$, 
cannot be expressed in a simple form in terms of $\rm T_{\rm AA}({b})$. It is of order unity for the most central
collisions ($b=0$), f$_2$/f$_1=$~4/3--16/15 for Gaussian and hard-sphere profiles respectively, 
but it is suppressed in comparison with the third leading term by an extra factor $\sim$2/A. 
For not very-peripheral collisions (f$_{\%}\approx$~0--65\%), the DPS cross section in a (thin) impact-parameter
range can be approximated by 
\begin{eqnarray} 
\sigmaDPS_{(AA\to a b)}[b_1,b_2] \approx \sigmaDPS_{(NN \to a b)}\cdot \sigmaeffpp \cdot
\rm f_{\%} \,\sigma_{\rm AA} \cdot \langle\rm T_{\rm AA}[b_1,b_2]\rangle^2 =
\left(\frac{m}{2}\right)\, \sigmaSPS_{(NN \to a)} \cdot \sigmaSPS_{(NN \to b)} \cdot
\rm f_{\%} \,\sigma_{\rm AA} \cdot \langle\rm T_{\rm AA}[b_1,b_2]\rangle^2\,.
\label{eq:doubleAA_b_final}
\end{eqnarray}
Taking the ratio of this expression over Eq.~(\ref{eq:singleAA_b}), one obtains the corresponding ratio of double to
single-parton-scattering cross sections as a function of impact-parameter:
\begin{eqnarray} 
(\sigmaDPS_{(AA\to a b)}/\sigmaSPS_{(AA \to a)})[b_1,b_2] \approx
\left(\frac{m}{2}\right) \, \sigmaSPS_{(NN \to b)} \cdot \langle\rm T_{\rm AA}[b_1,b_2]\rangle\,.
\label{eq:doubleAA_b_ratio}
\end{eqnarray}
This analytical expression neglects the first and second terms of Eq.~(\ref{eq:doubleAA_b}). In the centrality percentile
f$_{\%}\approx$~65--100\% the second term would add about 20\%  more DPS cross-sections, and for very
peripheral collisions (f$_{\%}\approx$~85--100\%, where $\langle\rm T_{\rm AA}[b_1,b_2]\rangle$ is of order or
less than $1/\sigmaeffpp$) the contributions from the first term are also non-negligible.\\

\paragraph*{\bf Results --} 
\label{sec:4}

\begin{figure}[htpb]
\epsfig{figure=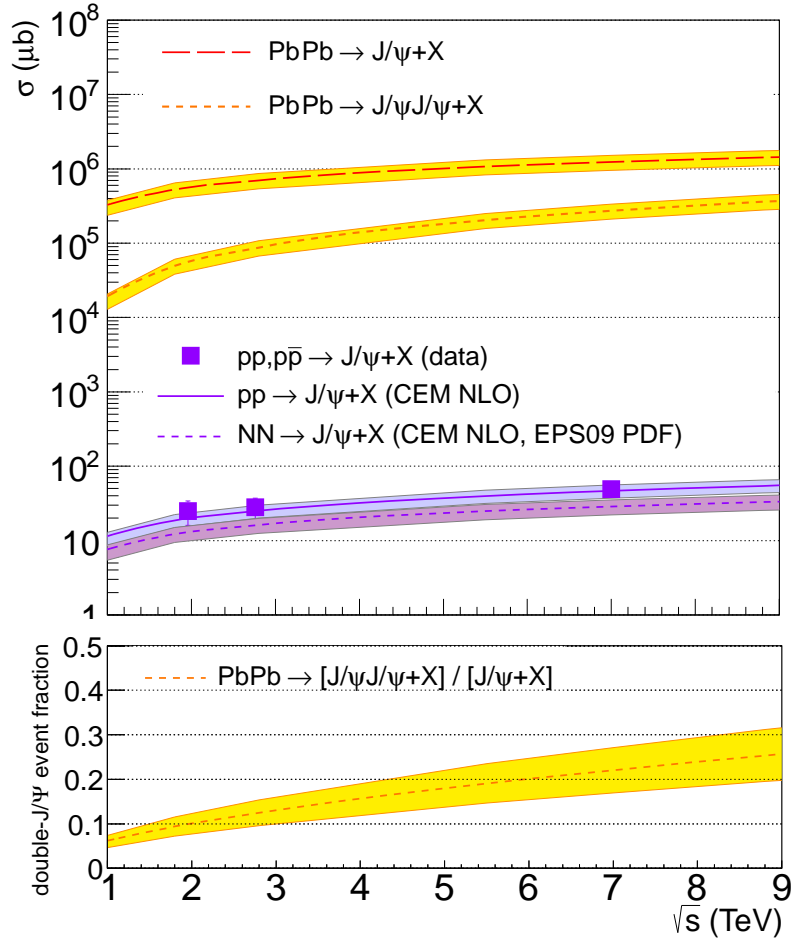,width=0.6\columnwidth}
\caption{Top: Cross sections for prompt-$\jpsi$ production in \pp, \NN, and \PbPb\ 
collisions and for double-parton $\jpsi\jpsi$ in \PbPb, as a function of c.m. energy. 
Bottom: Fraction of prompt-$\jpsi$ events where a pair of $\jpsi$ is produced in \PbPb\
collisions, as a function of c.m. energy. The bands show the nuclear PDF and scales uncertainties in quadrature.
\label{fig:sigmasqrts}}
\end{figure}

From Eqs.~(\ref{eq:sigmaAADPS}), with $m=1$, and (\ref{eq:sigmaeffAA}) we can compute the expected double-parton cross
sections for $\jpsi$-pair production in \PbPb\ from the single-parton $\jpsi$ cross sections in
nucleon-nucleon collisions, $\sigmaSPS_{(NN \to \jpsi\,X)}$, 
obtained with the color evaporation model (CEM)~\cite{Vogt:2010aa} cross-checked with the existing \pp\
and \ppbar\ data, 
and taking into account nuclear PDF modifications~\cite{Eskola:2009uj}.
The SPS cross sections for prompt $\jpsi$, after subtraction of the decay contributions from bottom
mesons, 
have been measured down to zero $p_T$ in \ppbar\ at
$\sqrts$~=~1.96~TeV at rapidities $|y|<$~0.6~\cite{Acosta:2004yw} and in \pp\ at $\sqrts$~=~2.76~TeV
($|y|<$~0.9~\cite{Abelev:2012kr}, 2~$<|y|<$~4.5~\cite{Aaij:2012ana}) and 7~TeV
($|y|<$~0.9~\cite{Abelev:2012gx}, 1.6~$<|y|<$~2.4~\cite{Khachatryan:2010yr}, 
2~$<|y|<$~4.5~\cite{Aaij:2011jh}). Empirical extrapolations to total $\jpsi$ cross sections at the LHC can be obtained
by integrating a Gaussian distribution fitted to the data points measured at different $y$. The Tevatron
midrapidity cross section can be extrapolated to full-rapidity with the prescription of 
\cite{Bossu:2011qe}. The values obtained, with their propagated uncertainties, are listed in
Table~\ref{tab:1} and shown as data points in Fig.~\ref{fig:sigmasqrts} (top) as a function of the c.m. energy.
Recent next-to-leading-order (NLO) CEM predictions for $\sigmaSPS_{(pp \to \jpsi\,X)}$
with theoretical scales $\mu_F$~=~1.5$\,m_c$ and $\mu_R$~=~1.5$\,m_c$ for a c-quark mass $m_c$~=~1.27~GeV
(solid curve)~\cite{Vogt:2012vr}, agree well with the experimental data including a $\pm$20\% uncertainty from
the scales. The corresponding values for $\sigmaSPS_{(NN \to \jpsi\,X)}$ are obtained from the CEM \pp\ cross
sections scaled by the EPS09 nuclear PDF shadowing. 
In the relevant (x,Q$^2$)~$\approx$~(10$^{-3}$, m$_{\jpsi}^2$) region, 
the Pb gluon PDF is moderately depleted, by a factor of $(1-S_{\rm g,Pb})\approx$~10\%--20\% with respect to the free
nucleon density, resulting in a reduction of the $g\,g \to \jpsi+X$ yields 
by a factor of $(1-S_{\rm g,Pb}^2)\approx$~20\%--35\% (dashed-dotted line in Fig.~\ref{fig:sigmasqrts}, top).
The EPS09 uncertainties, of the order of $\pm$10--15\%, 
have been propagated in quadrature with those associated with the theoretical scales,
into the \NN\ cross sections. 
We note that the EPS09 parametrization is clearly favored by the $\jpsi$ photoproduction
data measured by ALICE in ultraperipheral \PbPb\ collisions at 2.76~TeV~\cite{Abbas:2013oua}.\\ 

\begin{table}[htbp]
\caption{Total cross sections at LHC energies for the production of prompt $\jpsi$ in
single-parton-scatterings (SPS) in \pp, \NN, \PbPb\ collisions, and of prompt $\jpsi$-pairs in
double-parton-scatterings (DPS) in \PbPb. 
The \pp\ values are extrapolated from experimental data, the \NN\ values are a CEM prediction
including EPS09 nuclear PDFs, and the \PbPb\ results are derived 
from the \NN\ cross sections via the quoted equations.
\label{tab:1}}
\begin{center}
\begin{tabular}{llcccc}\hline
 Process  & Cross section   &  &  $\sqrtsnn$ & (TeV)  &  \\
          &  & 1.96 & 2.76 & 5.5 & 7.0 \\\hline
$\sigmaSPS\,$(\pp, \ppbar$\,\to \jpsi\,X$) [$\mu$b] & measured (extrapolated) & 25. $\pm$ 9. & 28. $\pm$ 8. & -- & 49. $\pm$ 9. \\
$\sigmaSPS\,$(\NN$\,\to \jpsi\,X$)  [$\mu$b] & CEM(NLO)+EPS09 PDF, Eq.~(\ref{eq:hardS}) & 14. $\pm$ 4. & 16. $\pm$ 3. & 25. $\pm$ 5. & 29. $\pm$ 6.\\
$\sigmaSPS\,$(\PbPb$\,\to \jpsi\,X$) [mb] &  Eq.~(\ref{eq:sigmaSPSAA}) & 600 $\pm$ 140 & 700 $\pm$ 150 & 1100 $\pm$ 250 & 1250 $\pm$ 280 \\
$\sigmaDPS\,$(\PbPb$\,\to \jpsi\jpsi\,X$) [mb] & Eqs.~(\ref{eq:sigmaAADPS})--(\ref{eq:sigmaeffAA}) & 65 $\pm$ 15 & 90 $\pm$ 20 & 200 $\pm$ 50 & 270 $\pm$ 60 \\\hline
\end{tabular}
\end{center}
\end{table}


The two uppermost curves in the top panel of Fig.~\ref{fig:sigmasqrts} show the resulting \PbPb\ cross sections for
single- and double-$\jpsi$ production, whereas their ratio is shown in the bottom panel. At the nominal \PbPb\
energy of 5.5~TeV, single prompt-$\jpsi$ cross sections (dashed curve) amount to about 1~b, and
$\sim$20\% of such collisions are actually accompanied by the production of a second $\jpsi$ from a
double parton interaction (dotted curve), whereas such processes are negligible at RHIC energies. 
The rise of the DPS/SPS ratio tends to slow down at higher $\sqrtsnn$
as the nuclear PDF shadowing (which enters squared in the numerator but only linearly 
in the denominator) increases, thereby reducing the total double-$\jpsi$ yields. 
The yellow bands in Fig.~\ref{fig:sigmasqrts}, amounting to about $\pm$25\%, include in quadrature the EPS09
PDF and theoretical scales uncertainties. 

\begin{figure}[htpb]
\centering
\epsfig{figure=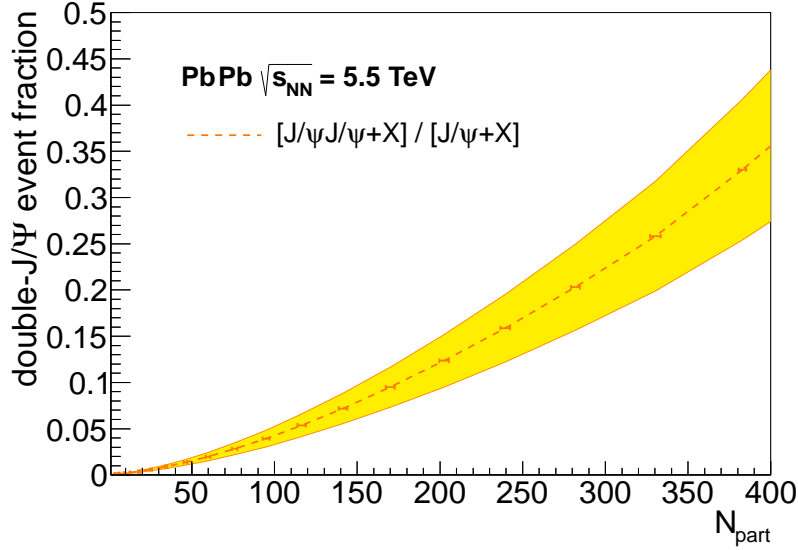,width=0.6\columnwidth}
\caption{Fraction of prompt-$\jpsi$ events in \PbPb\ collisions at 5.5~TeV where a $\jpsi$-pair is produced 
from double-parton scatterings as a function of the reaction centrality (given by N$_{\rm part}$), according
to Eq.~(\ref{eq:doubleAA_b_ratio}). The band shows the EPS09 PDF plus scale uncertainties.
\label{fig:SPSDPS_ratio_vs_centrality}}
\end{figure}

The ratio of single- to double-$\jpsi$ production, Eq.~(\ref{eq:doubleAA_b_ratio}), 
as a function of the reaction centrality quantified by the number of participant nucleons (0~$<$~N$_{\rm part}~<$~2\,A)
in \PbPb\ at 5.5~TeV is shown in Fig.~\ref{fig:SPSDPS_ratio_vs_centrality}. 
The probability of $\jpsi$-pair production increases rapidly 
and at the highest centralities (lowest impact parameters),
about 35\% of the \PbPb$\to\jpsi+X$ collisions have a second $\jpsi$ in the final state. 
We note that the DPS cross sections have to be understood as inclusive values, 
but they do not represent an extra contribution to the total prompt-$\jpsi$ rates since they are
already part of the \PbPb$\to\jpsi+X$ cross section.\\

Our predictions can be experimentally confirmed by measuring the cross sections for the simultaneous 
production of two $\jpsi$ mesons in the same \PbPb\ event via their visible dilepton decay channels. 
The two $\jpsi$ mesons issuing from double-parton scatterings have on average identical $p_T$ and $y$
distributions. 
At LHC energies, the cross section per unit-rapidity for single-$\jpsi$ amounts to 
$d\sigma_{\jpsi}/dy \approx \sigma_{\jpsi}/8$ at the (low-$p_T$) rapidities covered by ALICE (at $y=0$)
and CMS (at $y=2$), the detector acceptance and reconstruction efficiencies reduce the measured yield by factors
of 
$\sim$12--14~\cite{Suire:2012gt,Khachatryan:2010yr}, and the dilepton branching ratio amounts to about 6\%.  
Squaring all these quantities for the case of $\jpsi$-pair production results in a final
reduction factor of order 3$\cdot$10$^{-7}$ for both rapidity ranges. Thus, at 5.5~TeV one would 
expect a visible DPS cross section of about $d\sigmaDPSjpsijpsi/dy|_{y=0,2} \approx$~60~nb per
dilepton decay mode, i.e. about 240 double-$\jpsi$ events per unit-rapidity in the four
combinations of dielectron and dimuon channels in 1~nb$^{-1}$ of integrated luminosity, assuming no 
in-medium suppression (accounting for it would reduce the yields by a two-fold factor, see below). 
The same estimates for the 15~(150)~$\mu$b$^{-1}$ of \PbPb\ data already collected at 
2.76~TeV result in about 3.5~(35) double-$\jpsi$ events in ALICE (CMS) at mid (forward) rapidities. The
combinatorial background of dilepton pairs with invariant masses around m$_{\jpsi}$
needs to be taken into account in order to carry out such a measurement on an event-by-event basis.\\

\paragraph*{\bf Discussion and conclusions --}
\label{sec:5}
The \PbPb\ cross sections discussed so far include initial-state nuclear PDF modifications but no
final-state effects 
which can modify the final measured yields. Experimentally, \PbPb\ collisions at 2.76~TeV show a two-fold
reduction of the MB $\jpsi$ yields with respect to \pp, i.e. 
$R_{\rm AA}^{\rm MB}=\sigma_{\rm AA}/(A^2\cdot\sigma_{\rm pp})\approx$~0.5, whereas the corresponding 
value amounts to $R_{\rm AA}^{\rm MB}\approx$~0.4 at RHIC. In central \PbPb\ collisions, the $\jpsi$ yields
at the LHC are even less depleted  ($R_{\rm AA}^{\rm cent}\approx$~0.5) than at RHIC 
($R_{\rm AA}^{\rm cent}\approx$~0.2--0.3).
Assuming that the dominant suppression at both energies is due to the ``melting'' of the $\jpsi$ state in the QGP,
the smaller LHC suppression has been interpreted as indicative of a new component of regenerated $\jpsi$ from
\ccbar\ recombination~\cite{Andronic:2007bi,Zhao:2011cv}, accounting for up to 30\% of the
final production. Such an additional contribution has nothing to do with the primordial DPS
processes discussed here which, as aforementioned, are already accounted for in the total prompt-$\jpsi+X$
yields. In particular, suppression due to color deconfinement in the plasma should, on average, affect equally
the doubly-produced $\jpsi$'s and, thus, the overall $R_{\rm AA}$ suppression factor should remain the same
independently if the $\jpsi$'s are produced in the same or in two different \PbPb\ collisions. Reciprocally, our
results demonstrate that the observation of double (or higher-multiplicity) $\jpsi$ production in a given \PbPb\
event should not be wrongly interpreted as indicative of extra contributions from regenerated $\jpsi$'s, as
DPS processes are an intrinsic component of the total $\jpsi$ production with or without final-state QGP
effects. Such a standard assumption is quantitatively substantiated in this work for the first time.\\

In summary, we have derived a simple generic expression for double-parton-scattering (DPS) cross sections in heavy-ion
collisions as a function of the elementary single-parton cross sections in nucleon-nucleon collisions, and
an effective $\sigmaeffAA$ parameter dependent on the transverse profile of the system.
The DPS cross sections in \AaAa\ are found to be enhanced by a factor of $A^{3.3}/5$, to be compared with the
$A^2$-scaling of single-parton scatterings. We have studied the case of $\jpsi$-pair production at LHC
energies and found that DPS constitute an important fraction of the total prompt-$J/\psi$
cross sections, amounting to 20\% (35\%) of the primordial production in minimum-bias (most central) \PbPb\
collisions. At 5.5~TeV, about 240 double-$\jpsi$ events are expected per unit rapidity in the dilepton decay
channels (in the absence of final-state suppression) for an integrated luminosity of 1~nb$^{-1}$, providing a
quantitative test of the predictions presented here. Pair-production of pQCD probes issuing from
double-parton-scatterings represents an important feature of heavy-ion collisions at the LHC and needs to be
taken into account in any attempt to fully understand the event-by-event characteristics of any yield suppression
and/or enhancement observed in \PbPb\ compared to \pp\ data.\\

\paragraph*{Acknowledgments --}
We are grateful to T.~Dahms, J.Ph.~Lansberg, I.~Lokhtin, G.~Martinez, C.~Suire and H.~Woehri for useful discussions,
to R.~Vogt for providing the latest CEM predictions, and in particular to A.~Morsch for corrections to previous
versions of this paper as well as for valuable quantitative cross checks of the results presented here.
This work is partly supported by the CERN-RFBR Joint Research Grant No. 12-02-91505.



\begin{thebibliography}{99}

\bibitem{Brambilla:2010cs}N.~Brambilla {\it et al.}, Eur.\ Phys.\ J.\ C {\bf 71} (2011) 1534

\bibitem{Lansberg:2006dh}J.~P.~Lansberg, Int.\ J.\ Mod.\ Phys.\ A {\bf 21} (2006) 3857
\bibitem{matsui_satz}T. Matsui and H. Satz, Phys. Lett. B178 (1986) 416 
\bibitem{Datta:2003ww}S.~Datta, F.~Karsch, P.~Petreczky and I.~Wetzorke, Phys.\ Rev.\ D {\bf 69} (2004) 094507
\bibitem{Karsch:2005nk}F.~Karsch, D.~Kharzeev and H.~Satz, Phys.\ Lett.\ B {\bf 637} (2006) 75

\bibitem{Aad:2010aa}G.~Aad {\it et al.} [ATLAS], Phys.\ Lett.\ B {\bf 697} (2011) 294 
\bibitem{Chatrchyan:2012np}S.~Chatrchyan {\it et al.} [CMS], JHEP {\bf 1205} (2012) 063
\bibitem{Abelev:2012rv}B.~Abelev {\it et al.} [ALICE], Phys.\ Rev.\ Lett.\ {\bf 109} (2012) 072301
\bibitem{Suire:2012gt}C.~Suire [ALICE], arXiv:1208.5601 [hep-ex]; 
A.~Maire [ALICE], arXiv:1301.4058 [hep-ex]  
\bibitem{Adare:2006ns}A.~Adare {\it et al.} [PHENIX], Phys.\ Rev.\ Lett.\ {\bf 98} (2007) 232301;
Phys.\ Rev.\ C {\bf 84} (2011) 054912
\bibitem{Chatrchyan:2012mb} S.~Chatrchyan {\it et al.}  [CMS], Phys.\ Rev.\ Lett.\  {\bf 109} (2012) 152303

\bibitem{Andronic:2007bi}A.~Andronic, P.~Braun-Munzinger, K.~Redlich and J.~Stachel, Phys.\ Lett.\ B {\bf 652} (2007) 259
\bibitem{Zhao:2011cv}X.~Zhao and R.~Rapp, Nucl.\ Phys.\ A {\bf 859} (2011) 114

\bibitem{MPI}P.~Bartalini and L.~Fano (eds.) {\it et al.}, Proceeds. MPI'08, arXiv:1003.4220 [hep-ex]; 
Proceeds. MPI'11,
arXiv:1111.0469 [hep-ph] 

\bibitem{Abe:1997bp} F.~Abe {\it et al.} [CDF], Phys.\ Rev.\ Lett.\ {\bf 79} (1997) 584 
\bibitem{Abazov:2009gc}V.~M.~Abazov {\it et al.} [D0], Phys.\ Rev.\ D {\bf 81} (2010) 052012
\bibitem{Aad:2013bjm}G.~Aad {\it et al.} [ATLAS], arXiv:1301.6872 [hep-ex]; 
PAS-FSQ-12-028 [CMS]
\bibitem{Aaij:2011yc}R.~Aaij {\it et al.} [LHCb], Phys.\ Lett.\ B {\bf 707} (2012) 52
\bibitem{Abelev:2012rz}B.~Abelev {\it et al.} [ALICE], Phys.\ Lett.\ B {\bf 712} (2012) 165
\bibitem{Kom:2011bd} C.-H. Kom, A. Kulesza and W.J. Stirling, Phys. Rev. Lett. {\bf 107} (2011) 082002
\bibitem{Baranov:2011ch} S.P. Baranov, A.M. Snigirev and N.P. Zotov, Phys. Lett. B {\bf 705} (2011) 116
\bibitem{Novoselov:2011ff} A.A. Novoselov, arXiv:1106.2184 [hep-ph]
\bibitem{Baranov:2012re} S.P. Baranov, A.M. Snigirev, N.P. Zotov, A.Szczurek and W. Schafer, Phys.\ Rev.\ D {\bf 87} (2013) 034035

\bibitem{d'Enterria:2012qx}D.~d'Enterria and A.~M.~Snigirev, Phys.\ Lett.\ B {\bf 718} (2013) 1395
\bibitem{Diehl:2011yj}M.~Diehl, D.~Ostermeier and A.~Schafer, JHEP {\bf 1203} (2012) 089

\bibitem{d'Enterria:2003qs}D.~d'Enterria, nucl-ex/0302016 

\bibitem{Strikman:2001gz}M.~Strikman and D.~Treleani, Phys.\ Rev.\ Lett.\ {\bf 88} (2002) 031801

\bibitem{Treleani:2012zi} D.~Treleani and G.~Calucci, Phys.\ Rev.\ D {\bf 86} (2012) 036003 
\bibitem{deJager}C.W.~deJager, H.~deVries, and C.~deVries, Atomic Data and Nuclear Data Tables {\bf 14} (1974) 485

\bibitem{Lokhtin:2000wm}I.P. Lokhtin and A.M. Snigirev, Eur.\ Phys.\ J. \ C {\bf 16} (2000) 527

\bibitem{Vogt:2010aa}R.~Vogt, Phys.\ Rev.\ C {\bf 81} (2010) 044903

\bibitem{Eskola:2009uj}K.J.~Eskola, H.~Paukkunen and C.A.~Salgado, JHEP {\bf 0904} (2009) 065

\bibitem{Acosta:2004yw}D.~Acosta {\it et al.} [CDF], Phys.\ Rev.\ D {\bf 71} (2005) 032001

\bibitem{Abelev:2012kr}B.~Abelev {\it et al.} [ALICE], Phys.\ Lett.\ B {\bf 718} (2012) 295
\bibitem{Aaij:2012ana}R.~Aaij {\it et al.} [LHCb], JHEP {\bf 1302} (2013) 041 

\bibitem{Abelev:2012gx}B.~Abelev {\it et al.} [ALICE], JHEP {\bf 1211} (2012) 065
\bibitem{Khachatryan:2010yr} V.~Khachatryan {\it et al.} [CMS], Eur.\ Phys.\ J.\ C {\bf 71} (2011) 1575
\bibitem{Aaij:2011jh} R.~Aaij {\it et al.} [LHCb], Eur.\ Phys.\ J.\ C {\bf 71} (2011) 1645

\bibitem{Bossu:2011qe}F.~Bossu {\it et al.}, 
 arXiv:1103.2394 [nucl-ex] 
\bibitem{Vogt:2012vr}R.~Vogt, R.~E.~Nelson and A.~D.~Frawley, arXiv:1207.6812 [hep-ph] 

\bibitem{Abbas:2013oua} E.~Abbas {\it et al.}  [ALICE],
  arXiv:1305.1467 [nucl-ex]



\end{thebibliography}
\end{document}